\documentclass[sn-mathphys-num]{sn-jnl}

\usepackage{graphicx}%
\usepackage{multirow}%
\usepackage{amsmath,amssymb,amsfonts}%
\usepackage{amsthm}%
\usepackage{mathrsfs}%
\usepackage[title]{appendix}%
\usepackage{xcolor}%
\usepackage{textcomp}%
\usepackage{manyfoot}%
\usepackage{booktabs}%
\usepackage{algorithm}%
\usepackage{algorithmicx}%
\usepackage{algpseudocode}%
\usepackage{listings}%

\begin{document}

\title[Article Title]{Generally covariant evolution equations from a cognitive treatment of time}

\author{\fnm{Per} \sur{\"Ostborn}}

\affil{\orgdiv{Division of Mathematical Physics}, \orgname{Lund University}, \city{Lund}, \postcode{221 00}, \country{Sweden}}

\abstract{
The treatment of time in relativity does not conform to that in quantum theory. To resolve the discrepancy, a formalization of time is introduced in an accompanying paper, starting from the assumption that the treatment of time in physics must agree with our cognition. The formalization has two components: sequential time $n$ and relational time $t$. The evolution of physical states is described in terms of $n$. The role of $t$ is to quantify distances between events in space-time. There is a space-time associated with each $n$, in which $t$ represents the knowledge at time $n$ about temporal distances between present and past events. This approach leads to quantum evolution equations expressed in terms of a continuous evolution parameter $\sigma$, which interpolates between discrete sequential times $n$. Rather than describing the evolution of the world at large, these evolution equations provide probabilites of a set of predefined outcomes in well-defined experimental contexts. When the context is designed to measure spatio-temporal position $(x,t)$, time $t$ becomes an observable with Heisenberg uncertainty $\Delta t$ on the same footing as $x$. The corresponding evolution equation attains the same symmetric form as that suggested by Stueckelberg in 1941. When the context is such that the metric of space-time is measured, the corresponding evolution equation may be seen as an expression of quantum gravity. In short, the aim of this paper is to propose a coherent conceptual basis for the treatment of time in evolution equations, in so doing clarifying their meaning and domain of validity.}


\maketitle

\section{Introduction}
\label{intro}

In this paper, a generally covariant form for quantum mechanical evolution equations is motivated, starting from a formalisation of time introduced in an accompanying paper, where two aspects of time are identified and separated \cite{ostborn}. This formalisation is derived from an epistemic approach to physics, where the entire cognitive content of the concept of time is carried over to the physical model, and this cognitive content can be understood as a \emph{form of appearance} in the Kantian sense \cite{kant}.

At present, there are no generally accepted evolution equations that marry the general covariance of general relativity with the principles of quantum theory. One reason is that the treatment of time is fundamentally different in the two theories, a circumstance often referred to as \emph{the problem of time}. According to Karel Kucha\v{r} \cite{kuchar},

\begin{quote}
\emph{classical geometrodynamics does not seem to possess a natural time variable, while standard quantum theory relies quite heavily on a preferred time.}
\end{quote}
Therefore, the following question arises, in the words of Chris Isham \cite{isham}:

\begin{quote}
\emph{How should the notion of time be re-introduced into the quantum theory of gravity?}
\end{quote}
A possible answer is hinted at in this paper. It is organized as follows. In section \ref{noconform}, the incompatibility of the treatment of time in relativity and quantum theory is discussed, suggesting the need to reconsider the notion of time in physical models. This need is emphasized in section \ref{incoherent}, where it is argued that the treatment of time is conceptually incoherent in quantum theory itself. In order to remove this incoherence and to make quantum mechanical evolution equations more symmetrical in their spatial and temporal variables, Stueckelberg introduced in 1941 a wave equation with an evolution parameter $\sigma$ in addition to the variable $t$, as discussed in section \ref{attempts}. A formally similar two-fold notion of time emerges from the cognitive analysis in the accompanying paper \cite{ostborn}, as summarised in section \ref{cognitive}.

In order to use such a cognitive two-fold model of time in evolution equations in a consistent fashion, the same cognitive approach should be used to describe the experimental contexts in which these equations are supposed to apply. Such a description is provided in section \ref{expcon}. Alternative outcomes of the experiment in such contexts are discussed in section \ref{alter}. Using the cognitive model of time in conjunction with the notion of experimental contexts with predefined alternative outcomes, a generally covariant form of evolution equations that conform with Stueckelberg's wave equation is motivated in section \ref{empiricallaw}. Thanks to the philosophically well-defined starting point, the interpretation and domain of validity of these evolution equations are made clear. Such evolution equations can be used to describe experimental contexts in which trajectories of objects are measured, as well as contexts in which the metric of space-time is measured. In the latter case, the equation corresponds to a quantum mechanical evolution equation for space-time itself, as discussed om section \ref{evolspace}. As such, it may be seen as an expression of quantum gravity.

The treatment in this paper is general and conceptual. No quantitative calculations or predictions are attempted.

\section{Time in relativity and quantum theory}
\label{noconform}

The main difference between relativity and quantum theory when it comes to time is that the temporal and spatial variables are treated in the same way in relativity, but differently in quantum theory.

In the latter theory, time $t$ is treated as a precisely defined evolution parameter, just like in Newtonian mechanics. The variable $t$ is not an observable, since there is no associated Hermitean operator or Heisenberg uncertainty. In contrast, position $x$ is clearly an observable in that sense, to which an uncertainty $\Delta x$ can be associated.

In relativity, both $t$ and $x$ can be regarded as observables. This goes for special relativity as well as for general relativity. A careful discussion of this point is attempted, since the coordinates will be assigned such a role when modified quantum mechancial evolution equations that conform to the requirements of general relativity are motivated in section \ref{empiricallaw}.

The fact that the spatial and temporal components of the position $(x,t)$ play the same formal role in relativity is clear from the fact that Lorentz transformations between equally valid reference frames mix temporal and spatial coordinates:

\begin{equation}
\begin{array}{lll}
\left(\begin{array}{c}x\\t\end{array}\right) & \rightarrow & \left(\begin{array}{c}\tilde{x}(x,t)\\\tilde{t}(x,t)\end{array}\right).
\end{array}
\end{equation}
Such transformations have physical effects, since they are related to time dilation and length contraction. This gives physical significance to the coordinates in special relativity.

In general relativity, the crucial geometric quantity is the metric $g_{\mu\nu}(x,t)$, having the coordinates as arguments. The role of these coordinates is diminished since the principle of general covariance, put forward by Albert Einstein as an essential feature of general relativity, means that the defining field equation

\begin{equation}
R_{\mu\nu}(x,t)-\frac{1}{2}R(x,t)g_{\mu\nu}(x,t)+\Lambda g_{\mu\nu}(x,t)=\frac{8\pi G}{c^{4}}T_{\mu\nu}(x,t)
\label{efield}
\end{equation}
is invariant under an arbitrary differentiable and invertible change of coordinates $(x,t) \rightarrow(\tilde{x},\tilde{t})$, meaning that the form of the field equation remains the same:

\begin{equation}
\tilde{R}_{\mu\nu}(\tilde{x},\tilde{t})-\frac{1}{2}\tilde{R}(\tilde{x},\tilde{t})\tilde{g}_{\mu\nu}(\tilde{x},\tilde{t})+\Lambda \tilde{g}_{\mu\nu}(\tilde{x},\tilde{t})=\frac{8\pi G}{c^{4}}\tilde{T}_{\mu\nu}(\tilde{x},\tilde{t})
\label{tefield}
\end{equation}
Here, $R_{\mu\nu}(x,t)$ is the Ricci tensor, $R(x,t)$ is the Ricci scalar, $\Lambda$ is the cosmological constant, $G$ is the gravitational constant, $c$ is the speed of light, and $T_{\mu\nu}(x,t)$ is the stress-energy tensor.

The transformed metric $\tilde{g}_{\mu\nu}(\tilde{x},\tilde{t})$ describes the same curved four-dimensional manifold as $g_{\mu\nu}(x,t)$. In so doing, $g_{\mu\nu}(x,t)$ takes over the description of the objective aspects of geometry from the coordinates $(x,t)$. Because of the arbitrariness of the coordinate system used to express the field equation (\ref{efield}), it may be argued that the variables $x$ and $t$ lack any physical significance whatsoever in general relativity. Some consider them to be arbitrary labels. However, this is not quite true.

The distinction between spatial and temporal coordinates is upheld in a qualitative way in general relativity, giving each of them a physical aspect. In any coordinate system, a direction specified by these coordinates can be chosen at each location on the four-dimensional manifold such that different points along that direction correspond to temporally separated events. It is possible for an observer to travel in that direction and interpret these points as a series of events happening at different times at her own location. In a similar way, three independent directions can be chosen such that a set of points in the subspace defined by these directions correspond to spatially separated events. It is possible for an observer to choose a direction of travel such that she interpret any two such points as simultaneous events occurring at different locations.

This is a consequence of the Lorentzian signature $(-,+,+,+)$ of the metric $g_{\mu\nu}$. Put differently, the relation between a pair of points specified by coordinates is either time-like or space-like.

This discussion can be rephrased in terms of the local light cone centered at the origin of the coordinate system. There is always one vector expressed in this coordinate system that can be placed inside this light cone and could be called 'time', whereas there are three other linearly independent vectors that can be placed outside the light cone, all of which could commonly be described as 'spatial', since they share this common trait. In other words, the Lorentzian signature of the metric provides a guide for physically meaningful choices of coordinate systems. The physicality of the Lorentzian metric and a coordinate system suited to describe it become intertwined.

For the purposes of this paper, which tries to motivate physics from cognitive or epistemic principles, it is also important to note that the numerical values of $x$ and $t$ used in practice in general relativity can always be associated with measurements made by observers. These measurements may correspond to direct observations, where events along the worldline of the observer are assigned temporal coordinates $t$ by means of their coincidences with the tickings of a clock carried by this observer, whereas spatial coordinates $x$ are assigned to events by means of their coincidences with markings on a measuring rod carried by the same observer. Albert Einstein stressed the need for such coincidences in his discussion of coordinates in relation to general relativity \cite{einstein}.

Coordinates of events far from the observer may be inferred by other means than direct observations. In the accompanying paper, such deduced events are called \emph{quasievents} \cite{ostborn}. For example, the distance to and timing of a lightning strike can be deduced from the time lag between the flash and the thunder. Coincidences with clocks and rods carried by the observer along her worldline are still needed in such measurements, but these coincidences are just input in a deduction together with knowledge of physical law, such as the speed of sound. Including such inferred coordinate assignments of distant spatio-temporal events, a coordinate system for an extended region of space-time can be constructed empirically, into which events and trajectories are placed.  

Of course, such coordinate systems can and must be constructed regardless the theoretical setting in which spatiotemporal aspects of physics are studied empirically, be it a quantum mechanics or a general relativity. The use of a set of identical rods and clocks with the aim to define a cartesian coordinate system empirically by means of events measured to be equidistant by these tools is as essential in general relativity as in Newtonian mechanics or special relativity, since it is the failure to construct a universal such cartesian system that provides the empirical evidence that general relativity is needed in the first place. Here, identical clocks and rods mean stiff rods whose end points coincide when carried by the same observer, and clocks whose ticks coincide in time when carried by the same observer.

The failure to construct a universal cartesian coordinate system in general relativity is manifested by the fact that temporal and spatial intervals between a given pair of events as measured by two observers depend not only on their relative state of motion, but also on their positions in a gravitational field. That is, they may vary with the position of the observer in any single attempted Lorentz frame. The global or 'objective' significance of the coordinates assigned to these events is therefore ripped away. Nevertheless, they have local empirical significance for each observer, and it was argued above that this local significance is a prerequisite to do physics as we know it, regardless the theoretical framework.

This is closely related to Einstein's equivalence principle, which can be expressed as the possibility for each observer to construct a valid local cartesian coordinate system in which she describes herself as being at rest at the origin. That the system is 'valid' means that there is no experiment she can do that would force her to disqualify this system as an inertial reference frame. Such a coordinte system is therefore sufficient to describe the dynamics of objects in each small enough neighborhood in a curved space-time, as these objects are influenced by forces, including gravity. Further, it often provides the substrate for a useful approximation for the spatio-temporal dynamics in larger regions, as proven by the early success of Newtonian celestial mechanics.

The local validity of the empirically created coordinate systems used to do empirical science can be associated to the local validity of quantum mechanical evolution equations that emerges from the epistemic approach pursued here. The common root is that both coordinate systems and the laws of physics are centered around and created by an experimenter doing physics. The limited domain of validity of evolution equations in the present approach is further discussed in section \ref{empiricallaw}.

\section{Time in quantum theory is incoherent}
\label{incoherent}

\begin{figure}[tp]
\begin{center}
\includegraphics[width=80mm,clip=true]{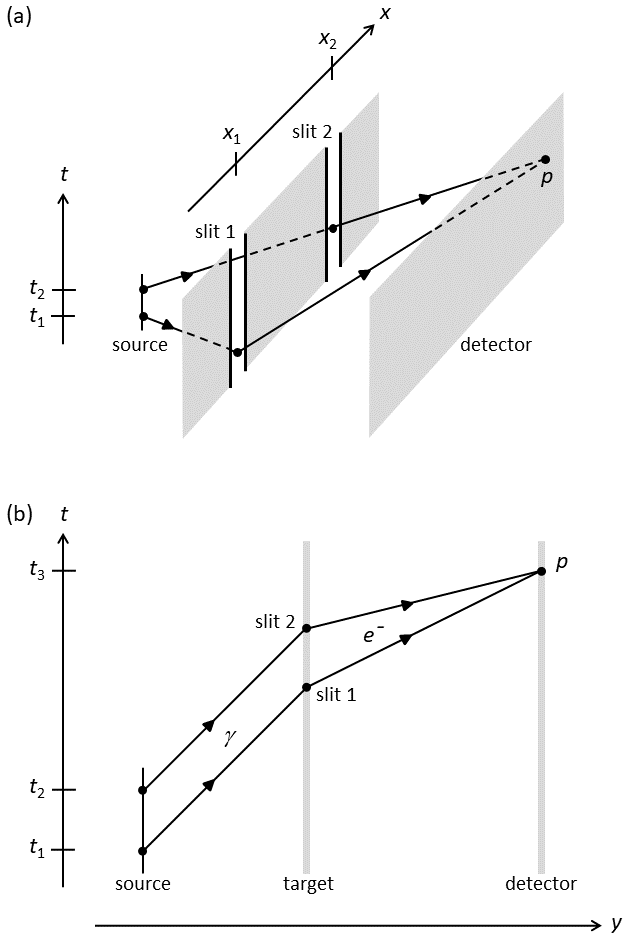}
\end{center}
\caption{(a) The double-slit experiment demonstrates not only interference of spatially separated paths, passing the slits at positions $x_{1}$ or $x_{2}$, but also interference of temporally separated paths, starting from the source at two different times $t_{1}$ and $t_{2}$. (b) Schematic illustration of a double-slit experiment with interference of paths with temporal separation only. Each of two laser pulses emitted at times $t_{1}$ and $t_{2}$ may ionize a single atom in a stationary target during two `temporal slits'. The two possibilities interfere when the emitted electron is detected.}
\label{Figure1}
\end{figure}

The currently used physical formalism is unsatisfactory not only in the sense that time is treated differently in quantum theory and general relativity, as discussed above, but also in the sense that the treatment of time is incoherent in quantum theory itself. The role of time in quantum theory is to evolve the state, and as such it is represented as a parameter with a well-defined value. To illustrate why this representation is insufficient, consider the double slit experiment with a vertical temporal axis $t$ added to the standard picture, as illustrated in Fig. \ref{Figure1}(a).

Assume 1) that a single object hits the detector screen at a point $p$ off the symmetry axis of the experimental setup, 2) that the speed of the object on its path from the source to the screen is known, and 3) that it is impossible to know which slit the object passes. Then there is interference between the two alternative paths. However, the two paths correspond to two different departure times from the source. Therefore, apart from the spatial interference between paths departing from the two slits located at positions $x_{1}$ and $x_{2}$, there is also temporal interference between paths departing from the source at times $t_{1}$ and $t_{2}$.

In other words, two possible but unobservable timings $t_{1}$ or $t_{2}$ of the event that the object is emitted from the source contributes to the probability that the object is detected at the point $p$ on the detector screen at a later time $t_{3}$. For the event corresponding to the object emission a temporal Heisenberg uncertainty $\Delta t\geq t_{2}-t_{1}$ must clearly be allowed, just as a spatial uncertainty $\Delta x\geq x_{2}-x_{1}$ is allowed for the event that the object passes a slit. Generally speaking, both temporal and spatial Heisenberg uncertainties $\Delta t$ and $\Delta x$ must be allowed in order to describe interference experiment in a satisfactory way.

This is not the case in the Schr\"odinger equation, where the same variable $t$ that measures timings of events is also used as a precisely defined evolution parameter, for which $\Delta t=0$ must be set. This fact suppresses the inherent symmetry between the spatial and temporal aspects of interference. To restore this symmetry, another kind of continuous evolution parameter $\sigma$ must be introduced to express differential evolution equations, thus releasing $t$ from this task and allowing it to display the desired uncertainty $\Delta t$, becoming an observable just like position $x$. 

Interference of two possible but unobservable timings of a single event has been demonstrated more explicitly in another kind of experiment, emphasizing this need \cite{wollenhaupt,lindner,ishikawa}. The basic idea behind these experiments is illustrated in Figure \ref{Figure1}(b). In the experiment by Wollenhaupt \emph{et al.} \cite{wollenhaupt} two ultra short laser pulses emitted at times $t_{1}$ and $t_{2}$ create a pair of temporal slits with separation $\Delta t=t_{2}-t_{1}$ at which an atom may be ionized, emitting an electron with a continuous energy spectrum. Interference between the two possible emission times of the single electron implies that the probability that an electron with given energy will be detected oscillates as a function of $\Delta t$. For a given time delay $\Delta t$ it also means that the probability to detect an electron oscillates as a function of its energy.

\section{Attempts to make quantum time coherent}
\label{attempts}

The idea to separate two aspects of time as one evolution parameter $\sigma$ and one observable time $t$ was explored already in 1941 by Ernst Stueckelberg \cite{stueckelberg1}. He parametrized trajectories in space-time according to $(x(\sigma),t(\sigma))$ in order to allow world lines that bend back and forth in time $t$ as $\sigma$ increases, thus modeling annihilation and pair production of particles and anti-particles. In 1942 Stueckelberg published a quantum mechanical version of his formalism \cite{stueckelberg2}, introducing a relativistic wave function $\Psi(x,t,\sigma)$ obeying an evolution equation $d\Psi/d\sigma=\mathcal{A}\Psi$, where the temporal component of the four-position $(x,t)$ is allowed to display Heisenberg uncertainty, just like the spatial components. In Stueckelberg's own words \cite{stueckelberg1}:

\begin{quote}
\emph{Le proc\'{e}d\'{e} de quantification de [Schr\"{o}dinger] peut alors \^{e}tre mis sous une forme o\`{u} l'espace et le temps interviennent d'une fa\c{c}on enti\`{e}rement sym\'{e}trique.}
\end{quote}

Several researchers have developed Stueckelberg's formalism further \cite{fanchi1,fanchi2,horwitz1,horwitz2,land}. However, there is still no generally accepted physical motivation for the introduction of the evolution parameter $\sigma$ in addition to $t$. Stueckelberg's original rationale is questionable on the grounds that a smooth particle trajectory $(x(\sigma),t(\sigma))$ that bends back and forth in time by necessity contains sections where it leaves its local light cone. If such bending is forbidden, on the other hand, it is possible to write $\sigma=\sigma(t)$, so that $\sigma$ loses its independent role and should be eliminated from the formalism.

Some physicists try to reserve an independent role for $\sigma$ as a moving \emph{now} inside a fixed space-time \cite{pavsic}. The parameter $\sigma$ is often related to the proper time along the world line of a given particle \cite{fanchi2,fanchi3}. A few researchers try to generalize this notion so that the same invariant $\sigma$ can describe the evolution of many particles, making it similar to the external, absolute time of Newtonian mechanics \cite{horwitz1,horwitz2,horwitz3,land}. Some authors argue that it is possible to construct a clock that measures the value of $\sigma$ \cite{fanchi3}, whereas others argue that $\sigma$ is not an observable \cite{horwitz3}. In any case, the parameter $\sigma$ must either be relativistically invariant, or play such a part in the formalism that relativistic transformations of spatio-temporal coordinates do not apply to it.

The picture of time introduced in this paper points to the same evolution equations as those proposed by Stueckelberg, with two temporal variables: one parameter $\sigma$ and one observable $t$. However, the interpretation of $\sigma$ will be different and hopefully more clear. The starting point is the cognitive model of time introduced in the accompanying paper \cite{ostborn}.

\section{A cognitive formalisation of time}
\label{cognitive}

This section summarizes the cognitive model of physical time offered in the accompanying paper \cite{ostborn}. The reader should consult that paper for more precise statements and more thourough explanations of the ideas involved.

The qualitative difference between temporal and spatial coordinates was discussed in section \ref{noconform}. In that regard, it may be noted that the trajectories of all massive objects are constrained to move within the local light cones in a given direction along the temporal axis, whereas trajectories may wiggle back and forth along the spatial axes. This feature corresponds to the flow of time, making it possible to order events along any given world line in a directed, linear sequence.

Such a temporal ordering applies locally for any given observer. It is argued in the accompanying paper that this ordering of events can be made global, applying to all events perceived by all observers regardless their location, without contradicting the relativity of simultaneity \cite{ostborn}. To this end, it is crucial to distinguish directly perceived events along the worldline of the observer from deduced quasievents, as defined in section \ref{noconform}. The temporal coordinates assigned to the latter are subject to the relativity of simultaneity, it is claimed, but not the temporal ordering of the former. The perceived events are fundamental from to the chosen cognitive perspective, whereas deduced events are not. The deduced properties of any such quasievent must have perceived events as input to the deduction. 

The proposed possibility to order all perceived events $e_{i}$ means that a universal temporal sequence of such events $\{\ldots,e_{i-1},e_{i},e_{i+1},\ldots\}$ can be defined. Two events in the sequence may be simultaneous, but the statement that no event $e_{i+1}$ occurs before $e_{i}$ is universally valid for all observers in this model.

A perceived event corresponds to a change of the state of knowledge of the perceiving observer, and thus to a change of the collective state of \emph{potential knowledge} $PK$ of all observers. Such a change defines an update $n\rightarrow n+1$ of \emph{sequential time} $n$.

The potential knowledge $PK(n)$ corresponds to the set of all proper interpretations and deductions about the world that can be made from perceptions at sequential time $n$, together with memories of previous such interpretations, derived from perceptions at earlier times. This knowledge corresponds to knowledge about a set of attributes $\{A,A',\ldots\}$ of potentially perceived objects $O$. The knowledge may be incomplete, meaning that the attribute value $a$ of $A$ is not precisely known, so that there is an entire set $\{a\}$ of such values that cannot be excluded given $PK(n)$. It is argued that $PK(n)$ is always incomplete in the sense that every value $a$ of every attribute $A$ of every object $O$ cannot be known at the same time $n$. This circumstance reflects the non-zero value of Planck's constant $h$ and Heisenberg's uncertainty relations.

It is possible to define an 'exact state' $Z$ in which the precise set of objects $O$ present in the world is given, together with precisely defined attribute values of each of these objects. The physical state $S(n)$ is defined as the set $\{Z\}$ of such states of the world that cannot be excluded given $PK(n)$. The assumed incompleteness of potential knowledge $PK(n)$ implies that $S(n)$ contains several elements $Z$ at all times $n$.

Physical law can be specified in part by an evolution operator $u_{1}$ such that

\begin{equation}
S(n+1)\subseteq u_{1}S(n),
\label{evop}
\end{equation}
where $u_{1}S_{n}\subset\mathcal{S}$ is the smallest set that guarantees that relation (\ref{evop}) is fulfilled, and $\mathcal{S}$ is the space of all exact states $Z$.

The physical state $S(n)$ describes the world as a whole. When doing physics, interest is most often restricted to the laws governing a particular object $O$ that is a proper subset of the world, even though this object may be large and contain many other objects as parts.  The state $S_{OO}(n)$ of such an object of study is specified by the set $\{Z_{O}\}$ of exact states $Z_{O}$ of the object that cannot be excluded given $PK(n)$. For two subjectively distinguishable objects $O$ and $O'$ it holds that $S_{OO}(n)\cap S_{O'O'}(n)=\varnothing$. Similarly, for two distinguishable subsequent states $S_{OO}(n)$ and $S_{OO}(n')$ of the same same object it holds that $S_{OO}(n)\cap S_{OO}(n')=\varnothing$.

It may happen that new knowledge is gained about $O$ at some seqential time $n+1$ that could not be predicted beforehand. Such an event can be formalised as

\begin{equation}
u_{O1}S_{OO}(n)\rightarrow S_{OO}(n+1)\subset u_{O1}S_{OO}(n),
\label{sored}
\end{equation}
where the object evolution operator $u_{O1}$ represents everything that can actually be predicted about $S_{OO}(n+1)$ at time $n$:

\begin{equation}
S_{OO}(n+1)\subseteq u_{O1}S_{OO}(n),
\label{oevop}
\end{equation}
in the same way as for the evolution operator $u_{1}$ in Eq. (\ref{evop}). The event (\ref{sored}) may be called a \emph{state reduction}, and corresponds to a wave function collapse in quantum mechanical language.

\begin{figure}[tp]
\begin{center}
\includegraphics[width=80mm,clip=true]{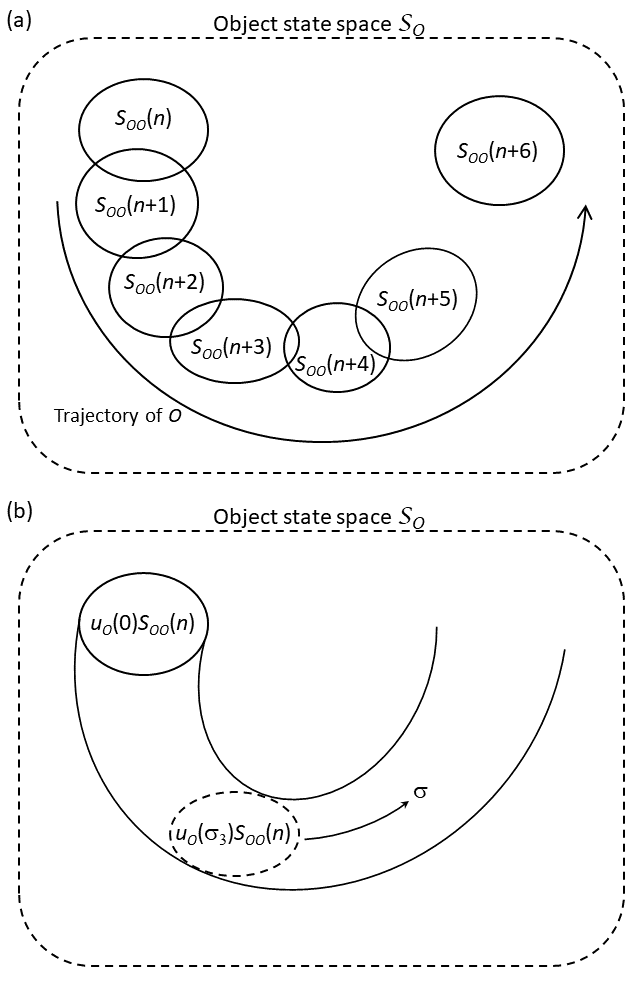}
\end{center}
\caption{(a) Object states $S_{OO}$ typically consist of several 'exact states' in state space $\mathcal{S}_{O}$ due to the incompleteness of potential knowledge $PK$. As a consequence, subsequent states may overlap, meaning that the object may not perceivably change from one sequential time $n$ to the next. (b) Such a seamless evolution of states suggests the use of a continuous evolution parameter $\sigma$.}
\label{Figure2}
\end{figure}

Consider the evolution of the object state $S_{OO}$ in the space $\mathcal{S}_{O}$ of all exact object states, as shown in Fig. \ref{Figure2}(a). Pairs of subsequent object states may overlap during an extended period of sequential time until a distinct change occurs, which happens at time $n+6$ in Fig. \ref{Figure2}(a). The overlapping states make the evolution smooth, in a sense, which invites the use of a continuous evolution parameter $\sigma$, as shown in Fig. \ref{Figure2}(b). Here, $u_{O}(0)\equiv I$ and $u_{0}(\sigma_{3})\equiv u_{O1}u_{O1}u_{O1}\equiv u_{O3}$.

It is crucial to note that the numerical value of $\sigma$ has no physical meaning. Any invertible change of variables $\tilde{\sigma}=f(\sigma)$ produces an equally valid evolution parameter $\tilde{\sigma}$. The only role of $\sigma$ is to parametrize the directed temporal sequence of object states continuously, like threading a series of pearls into a necklace. 

The potential knowledge $PK(n)$ at sequential time $n$ contains knowledge about past times $n-m$ in the form of memories, or knowledge that some perceptions at time $n$ can be properly interpreted as records about past times. Similarly, perceptions at time $n$ allows deductions about the future given physical law. This means that $PK(n)$, and the corresponding epistemic physical state $S(n)$, can be described as temporally holistic even though it is always anchored at a specific time $n$ that may be called \emph{now}.

In particular, each present time $n$ contains a map that specifies the spatio-temporal locations of objects in the past, at present, and in the future. This map is always blurry to some extent, given the assumed incompleteness of potential knowledge. The blurriness may stem from imprecise perceptions at present, from memories that are potentially lost, or from a lack of determinism that prevents us from making precise predictions.

\begin{figure}[tp]
\begin{center}
\includegraphics[width=80mm,clip=true]{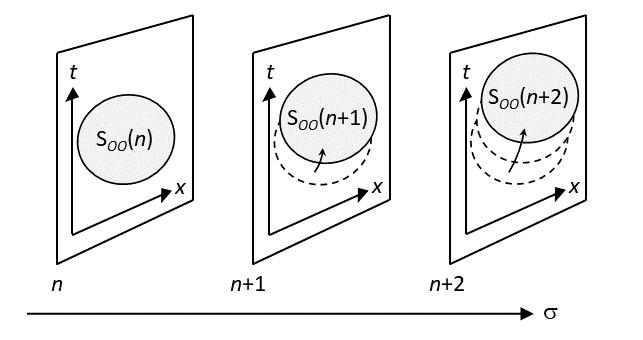}
\end{center}
\caption{The relation between the evolution parameter $\sigma$, sequential time $n$, and relational time $t$. There is an entire space-time spanned by the spatial axis $x$ and the temporal axis $t$ associated with each sequential time $n$, onto which the state $S_{OO}(n)$ of object $O$ can be projected. The seamless evolution of $S_{OO}(n)$ as $n$ increases can be described using the continuous parameter $\sigma$. At each time $n$ there are memories of past states of $O$, which are shown as states with dashed boundaries. They make it possible to define a trajectory $x(t)$ of $O$ at each time $n$, as indicated by curved arrows.}
\label{Figure3}
\end{figure}

As a temporal coordinate in the map of space-time defined at sequential time $n$, a new temporal variable $t$ is introduced, called \emph{relational time}. The coordinate $t$ in the spatio-temporal map becomes an observable, just like the corresponding spatial coordinate $x$. That the map is blurry means that the potential knowledge about the position $(x,t)$ of some object $O$ is incomplete at time $n$, corresponding to an object state $S_{OO}(n)$ containing several elements $Z_{O}$ or a Heisenberg uncertainty $(\Delta x, \Delta t)$. The situation is illustrated in Fig. \ref{Figure3}.

It should be emphasized that the measured value $t$ of relational time does not necessarily correspond to the present time $n$ at which this measurement is made. It may refer back to an event relating to a previous state of an object, occurring at some sequential time $n-m$. For example, astronomical events such as faraway supernovas that are observed \emph{now} must be assigned a relational time $t$ that refer to a distant past, possibly millions of years ago.

The idea to expand the fundamental physical description of the \emph{now} so that it contains information not only about the state of the world at that present time $n$, but also partial information about the past, is justifiable if and only if a strict epistemic approach to physics is adopted, in which the present physical state of the world corresponds to the present knowledge about the world. That knowledge contains information about the past via memories and records. In contrast, in a realistic model of the world, the information about the past in the form of memories and records is just a function of the present physical state of the brain. The only relevant aspect of the records will be their present physical state. Thus, memories and records become secondary from a realistic point of view, and the past does not become a necessary part of the description of the present.

From the subjective point of view, the expanded notion of the \emph{now} is the more natural one. Suppose that you listen to music. The appreciation of harmonies, and the emotional response they give rise to in the present, depend crucially on memories of sounds in the immediate past, to the extent that the music would cease to exist without these memories. That is, each present state of the listener contains both the present and the past in a crucial way; each fleeting \emph{now} encoded by $n$ can be unfolded to an entire temporal axis $t$. At the formal level of physical description, the very perception of a sound at a given moment relies on sensory recording during an extended period of time, since such a temporal interval is needed to define the frequencies that determine the sound that we hear at a given moment.

\section{Experimental contexts}
\label{expcon}

The evolution operator $u_{O}(\sigma)$ encapsulates what can be known at a given sequential time $n$ about the evolution of object $O$ at a later time $n+m$. Consequently, the expression $u_{O}(\sigma)S_{OO}$ has no direct relation to an actual observation of $O$ at such a later time, or a measurement of any of its attributes. In particular, the claimed incompleteness of knowledge means that state reductions must be allowed at any time $n+m$ according to Eq. (\ref{sored}), at which a subject learns something new about the object $O$ that could not have been predicted at time $n$.

The epistemic approach used in this paper dictates that any meaningful evolution equation should take such state reductions into account. Further, from the same perspective, such equations must correspond to well-defined experimental settings in which the values of attributes of the object of interest are measured. An evolution equation purporting to describe the dynamics of objects in and of themselves, without any relation to the means by which this dynamics is observed, has no physical meaning from the present point of view. This perspective is the same as that expressed by Niels Bohr \cite{bohr}:

\begin{quote}
\emph{I advocated the application of the word phenomenon exclusively to refer to the observations obtained under specified circumstances, including an account of the whole experimental arrangement.}
\end{quote} 

Such a well-defined setting may be called an \emph{experimental context} $C$. The observed object $O$ and the body of an aware observer $OB$ are both necessary components in $C$, according the epistemic premise of this paper. The observed object $O$ may be divided into a \emph{specimen} $OS$ and an \emph{apparatus} $OA$, with which values of attributes of the specimen of interest are measured. For example, the apparatus may be a rod or a clock, according to the relativistic discussion in section \ref{noconform}.

The concept of \emph{quasievents} was introduced in section \ref{noconform} as events that are not directly observed, so that their properties, such as the spatio-temporal location, must be calculated from actual observations. Such a quasievent can be associated with a corresponding deduced \emph{quasiobject}. When the specimen $OS$ is a quasiobject, the apparatus can be divided into a \emph{machine} $OM$ and a \emph{detector} $OD$, where changes in the state of the detector perceived by observer $OB$ define the outcome of the experiment. In such a context $C$, there has to be prior knowledge that correlates the readings of the detector with attributes of the specimen $OS$.

In the double-slit experiment shown in Fig. \ref{Figure4}, the specimen $OS$ is a particle ejected from a gun. This gun, together with the slits and supportive structure constitutes the machine $OM$. The detector $OD$ is the screen on which the particle ends up.

The state $S_{OO}(n_{i})$ of the observed object $O$ in a context $C$ is assumed to be prepared at an initial time $n_{i}$ so that there will be a definitive outcome of the experiment at some later final time $n_{f}$, given that the observer $OB$ does not interfere with the experiment after time $n_{i}$. This fact is part of potential knowledge $PK(n_{i})$. Figuratively speaking, the observer pushes a button at time $n_{i}$, and the experiment runs by itself until it reveals the outcome at some finite time $n_{f}$.

\section{Alternatives}
\label{alter}

The existence of state reductions according to Eq. (\ref{sored}) implies that different outcomes must be allowed for the same initial physical state $S_{OO}(n_{i})$ of object $O$ in an experimental context $C$. Therefore, the most precise prediction that an evolution equation can possibly make is to assign probabilities to all possible outcomes, given $S_{OO}(n_{i})$. In a proper context $C$, the outcome is assumed to be known at the start of the experiment at time $n_{i}$ to be one of a set of \emph{alternatives} $\{A_{j}\}$ relating to the value $a$ of a specific attribute $A$ of the specimen $OS$.

If the set of possible values $a$ of $A$ is finite, such as the two possible spin angular momentum values $\{a_{1},a_{2}\}$ of an electron along a given axis, then it is possible to identify an alternative with an individual attribute value according to $A_{j}=a_{j}$. More generally, the alternative $A_{j}$ corresponds to a set of values $\{a\}$ that cannot be excluded after the completion of the experiment a time $n_{f}$. If this set has more than one element, the potential knowledge $PK_{A}(n_{f})$ of the attribute $A$ of $OS$ is greater than $PK_{A}(n_{i})$ thanks to the experimental observation, but is still incomplete. This is clearly so whenever the possible values $a$ form a continuous set. For example, in the double-slit experiment in Fig. \ref{Figure4} the possible outcomes form a set of alternative positions $\{X_{j}\}$ that can be resolved on the detector screen, where each alternative $X_{j}$ corresponds to a continuum of exact positions $x$.

An experimental context $C$ is assumed to be such that the set of alternatives $\{A{j}\}$ known at time $n_{i}$ is \emph{complete} in the following sense. Let $S_{A}(n_{i})$ be the set of values $a$ of attribute $A$ of the specimen $OS$ that cannot be excluded when the experiment starts at time $n_{i}$. Then

\begin{equation}
\begin{array}{c}
A{j}\subset S_{A}(n_{i})\\
\bigcup A{j}=S_{A}(n_{i})\\
A_{j}\cap A_{j'}=\varnothing.
\end{array}
\end{equation}

Any complete set of alternatives is assumed to be finite, since the resolution power of a detector is always finite, and it always has finite size.

An entire set of attributes $\{A,A',\ldots\}$ can be measured within the same context $C$, corresponding to a set of alternatives $\{A_{j},A_{j'}',\ldots\}_{jj'\ldots}$. Simple examples are the spatial positions of different parts of a composite specimen, the four coordinates of the spatio-temporal position of the specimen, or a set of locations of the specimen that define its trajectory.

\section{Evolution equations}
\label{empiricallaw}

A set of probabilities $\{p(A_{j}),p(A_{j'}'),\ldots\}_{jj'\ldots}$ can be associated to complete sets of alternatives according to the above. They may be expressed as a function $P(a,a',\ldots)$ that is defined within the context $C$, and is determined by the experimental arrangement:

\begin{equation}
C\Rightarrow P(a,a',\ldots).
\label{cprob}
\end{equation}
Here, the domain of each argument $a$ is thought of as the corresponding set of alternatives $\{A_{j}\}$.

The function $P$ may be computed from a more fundamental entity, such as a contextually defined complex wave function $\Psi(a,a',\ldots)$. It is argued in Ref. \cite{ostborn2} that probabilities indeed need to be represented as complex numbers and computed via Born's rule, in order to make such an algebraic representation generally applicable to any context $C$.

To be able to say that physical law is understood it should be possible to describe how the probabilities of all conceivable outcomes of an observation depends on the variables that specify the context $C$ in which the observation takes place. However, these variables are observable attributes, the values of which cannot be assumed to be precisely known within context, due to the assumed incompleteness of potential knowledge.

This fact raises an obstacle in the chosen epistemic approach to physics. According to the principle of \emph{ontological minimalism} introduced in the accompanying paper \cite{ostborn}, physical law cannot be properly expressed by assuming more precise knowledge about these attribute values than can ever be obtained. Therefore, these values cannot be used as arguments in a function that specify probabilities of outcomes in the given experimental context. In particular, this excludes the observed relational time $t$ as such an argument. 

Consider instead the expression

\begin{equation}
S_{OS}(n_{f}-1)=f(S_{OS}(n_{i}),S_{OA}(n_{i})),
\label{contextf}
\end{equation}
where $S_{OS}(n_{f}-1)$ is the state of the specimen just before the observation that defines the outcome of the experiment, and $S_{OS}(n_{i})$ and $S_{OA}(n_{i})$ are the initial state of the specimen and apparatus, respectively. This expression is valid as long as the context $C$ is sufficiently insulated from the environment during the course of the experiment.  

Clearly, the duration of the experiment specified by $n_{f}-n_{i}$ depends on $S_{OS}(n_{i})$ and $S_{OA}(n_{i})$. If the outcome of an experiment is to be expressed explicitly as a function of a temporal variable, in a way that is correct as a matter of principle, then a precisely defined variable that is not an observable attribute should be used. This variable must nevertheless have physical meaning. The evolution parameter $\sigma$ fits this description. However, a careful discussion about its possible role in such evolution equations is needed.

\begin{figure}[tp]
\begin{center}
\includegraphics[width=80mm,clip=true]{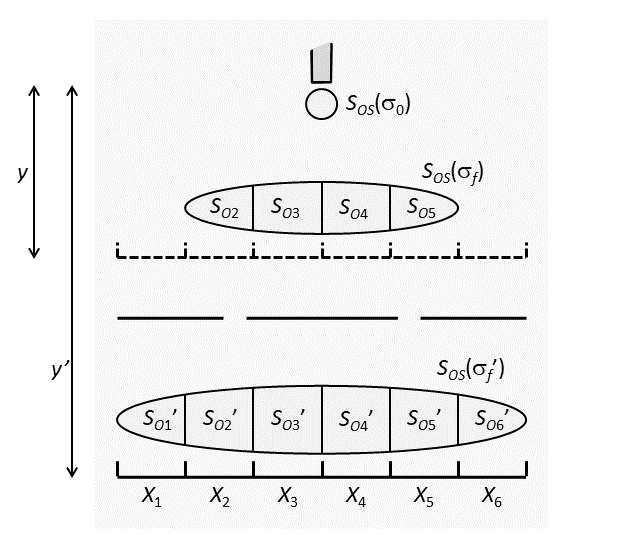}
\end{center}
\caption{Two members of a family $C(\sigma_{f})$ of experimental contexts $C$, where $\sigma_{f}$ is the value of the evolution parameter $\sigma$ just before detection of the particle in a double-slit experiment at time $n_{f}$ after its ejection from a gun at time $n_{i}$. A change of $\sigma_{f}$ corresponds in this family to a change of the position $y$ of the detector screen. The attribute $A$ that is measured in $C(\sigma_{f})$ is the position $X$ of the particle along the screen. The possible outcomes correspond to a set of alternatives $X_{j}$, in which each element in turn correspond to non-overlapping bins of values $\{x\}_{j}$. These sets are assumed to be the same in all contexts that belong to $C(\sigma_{f})$. The set $S_{Oj}$ corresponds to the object state $S_{OS}$ of the specimen given that alternative $X_{j}$ comes true.}
\label{Figure4}
\end{figure}

It is often possible to find an attribute $A$ specifying the context $C$ such that the time $n_{f}-n_{i}$ passed before the outcome of the experiment reveals itself is known to increase monotonically with its value $a$. One example is the distance $y$ between the particle gun and the detector screen in the double-slit experiment shown in Fig. \ref{Figure4}.

Let $\sigma_{f}$ be the final value of $\sigma$ at time $n_{f}-1$ just before the observation that defines the update to time $n_{f}$. Then $d\sigma_{f}/da>0$, and given that the variation of $A$ is the only variation of the initial state $S_{OA}(n)$ of the experimental apparatus $OA$ it is possible to write  

\begin{equation}
S_{OS}(n_{f}-1)=f_{\sigma}(S_{OS}(n_{i}),\sigma_{f}).
\label{sigmalaw}
\end{equation}

Different values of the argument $\sigma_{f}$ in Eq. (\ref{sigmalaw}) clearly corresponds to different experimental contexts $C$. A one-parameter family of contexts $C(\sigma_{f})$ is obtained. To make the notation simpler the subscript on $\sigma$ is dropped in the following, writing $C(\sigma)$. The meaning of the argument should be understood according to the preceding discussion.   

It was argued in Section \ref{cognitive} that the numerical value of $\sigma$ has no physical meaning in the sense that any invertible change of variables produces an equally valid evolution parameter. Since sequential time $n$ is updated at every event potentially perceived by anybody anywhere in the world, the numerical value $n_{f}-n_{i}$ is unknowable to any individual observer. For these two reasons, it is impossible to determine the parametrized physical law expressed in Eq. (\ref{sigmalaw}) empirically. However, a series of experiments can be performed in which the value $a$ of the associated observable attribute $A$ is changed. The set of such experiments corresponds to an empirical family of contexts $C(a)$ that can be employed to estimate physical law, as expressed in Eq. (\ref{sigmalaw}). Nevertheless, the parameter $\sigma$ cannot simply be replaced by the value $a$ of an observable as an argument in Eq. (\ref{sigmalaw}), as discussed above.

As a matter of principle, physical law transcends the present time $n$. However, to learn about it there is no choice but to use as tools the perceivable attributes at time $n$, such as the observed distance $y$ between the particle gun and the detector screen in Fig. \ref{Figure4} at the start of the experiment, together with the memories or records at that time of previous choices of $y$. To put it more succinct: there is physical law that transcends the empirical evidence, but it cannot be learned by transcending the empirical evidence.

Allowing for the entire family $C(\sigma)$ of experimental contexts, the contextual probability function $P$ in Eq. (\ref{cprob}) can be expressed as 

\begin{equation}
C(\sigma)\Rightarrow P(a,a',\ldots,\sigma).
\label{csprob}
\end{equation}

Since the evolution parameter $\sigma$ is continuous by definition, and the evolution can be seen as gradual according to the discussion in section \ref{cognitive}, a contextual evolution equation may be written as

\begin{equation}
\frac{dP}{d\sigma}=h(a,a',\ldots,\sigma).
\label{dprob}
\end{equation}

As noted above, the probability function $P$ is often conveniently computed from a wave function $\Psi$. It follows that the most convenient form of the evolution equation is often

\begin{equation}
\frac{d\Psi}{d\sigma}=\mathcal{A}\Psi(a,a',\ldots,\sigma),
\label{cspsi}
\end{equation}
for some evolution operator $\mathcal{A}$.

In order to highlight the different roles played by the two aspects of time employed in this paper, consider a family of experimental contexts in which a spatio-temporal position $(x,t)$ of the specimen is measured. In this case, the contextual evolution equation reads

\begin{equation}
\frac{d\Psi}{d\sigma}=\mathcal{A}_{xt}\Psi(x,t,\sigma).
\label{rtpsi}
\end{equation}
This is formally the same evolution equation as that suggested by Stueckelberg, as discussed in section \ref{attempts}. Equation (\ref{rtpsi}) is clearly generally covariant, that is, its form is invariant under any invertible and differentiable coordinate transformation $(x,t)\rightarrow (\tilde{x},\tilde{t})$:

\begin{equation}
\frac{d\tilde{\Psi}}{d\sigma}=\tilde{\mathcal{A}}_{xt}\tilde{\Psi}(\tilde{x},\tilde{t},\sigma).
\label{trtpsi}
\end{equation}
The transformed evolution operator $\tilde{\mathcal{A}}_{xt}$ should retain the same physical meaning as $\mathcal{A}_{xt}$. It will be argued in forthcoming papers that it can be related to the rest mass of the specimen $OS$, in a similar manner as the corresponding evolution operator in the ordinary Schr\"odinger equation is related to its energy.

The values of $(x,t)$ measured in the family of contexts $C(\sigma)$ may correspond either to a historical or to a present state of the specimen $OS$. A historical measurement may, for example, correspond to the location and timing of an astronomical event, such as a supernova. In that case the distant exploding star is a specimen $OS$ that can be categorized as a deduced quasiobject.

A measurement of the present spatio-temporal position may, on the other hand, correspond to an experimental context such as the double-slit experiment in Fig. \ref{Figure4}, in which the screen consists of an array of detectors akin to Geiger counters, which click whenever a particle is detected. The click of the counter at position $X_{j}$ at time $n_{f}$ prompts the immediate reading of a clock which provides an estimate $T_{j'}$ of the temporal coordinate $t$ of this event. The clock may be a stopwatch carried by the experimenter started at time $n_{i}$ and stopped at time $n_{f}$, so that the time of flight of the particle from the gun to the detector is measured in an reference frame at rest relative to the experimental setup $C$. 

Equation (\ref{rtpsi}) has the same form as Stueckelberg's evolution equation discussed in Section \ref{attempts}. However, the evolution parameter $\sigma$ is given a different role than in previous attempts to motivate Stueckelberg's equation. Here, it is considered to be closely associated to the flow of sequential time $n$. Since it is argued that $n$ corresponds to a universal directed ordering of events with no associated observable metric to describe temporal distances, $\sigma$ is not affected by Lorentz transformations between the reference frames of different observers. The physical meaning of this evolution parameter lies in the fact that it changes monotonically with $n$. Its continuity reflects the gradual change of identifiable objects or quasiobjects, as illustrated in Fig. \ref{Figure2}. The numerical value is arbitrary, and does not correspond to any observable attribute. It was noted in section \ref{attempts} that an evolution parameter $\sigma$ used to express Eq. (\ref{rtpsi}) should either be relativistically invariant, or play such a part in the formalism that relativistic transformations of spatio-temporal coordinates do not apply to it. The cognitive framework used in this paper to motivate the introduction of $\sigma$ clearly uses to the second option.

Nevertheless, in a given reference frame it is possible to introduce a \emph{natural parametrization} for families of contexts $C(\sigma)$ in which the present temporal coordinate $t$ of an event relating to the specimen $OS$ is measured. Such a natural parametrization is defined by the condition $d\sigma/d\langle t\rangle=1$, where $\langle t\rangle$ is the \emph{expected} relational time to be measured at sequential time $n_{f}$. This expected time is often known at the start of the experiment at time $n_{i}$ as a function of the experimental setup. In that case it is possible to rewrite Eq. (\ref{rtpsi}) as 

\begin{equation}
\frac{d\Psi}{d\langle t\rangle}=\mathcal{A}_{xt}\Psi(x,t,\langle t\rangle).
\end{equation}

Put in this way, the distinction between the two temporal variables in the evolution equation becomes subtle. This may be the root of the conflation of $\sigma$ and $t$ that has led to difficulties understanding the different roles played by the temporal variables used in quantum theory and relativity, and to confusion about how they should be made to fit together in a more general theory.

\section{Evolving space-times}
\label{evolspace}

The set of attributes $\{A,A',\ldots\}$ measured within the family of contexts $C(\sigma)$ described by the evolution equation (\ref{cspsi}) is arbitrary. To illuminate the discussion in section \ref{noconform} about geometry and coordinates it may be worthwhile to let $A$ be the metric $g_{\mu\nu}(x,t)$ at some point $(x,t)$, rather than letting $A$ correspond to these coordinates themselves. Assuming that it is appropriate to represent the probability function $P$ in Eqs. (\ref{csprob}) and (\ref{dprob}) by a complex-valued wave function via Born's rule regardless the nature of the attribute $A$, as suggested by the analysis in Ref. \cite{ostborn2}, the evolution equation for the metric takes the form

\begin{equation}
\frac{d\Psi}{d\sigma}=\mathcal{A}_{g}\Psi(g_{\mu\nu}(x,t),\sigma).
\label{gpsi}
\end{equation}
This equation is a neat and general way to express a situation in which the metric in an extended region of space-time is estimated by means of measurements at a finite number of locations $\{P_{1},P_{2},\ldots\}$. These locations are assumed to remain the same in the entire context family $C(\sigma)$ in the sense that the set of reference objects $\{O_{1},O_{2},\ldots\}$ that are necessary to define the locations remain the same. The evolution equation (\ref{gpsi}) can therefore be more explicitly, but less elegantly, expressed as 

\begin{equation}
\frac{d\Psi}{d\sigma}=\mathcal{A}_{g}\Psi(\left(g_{\mu\nu}\right)_{1},\left(g_{\mu\nu}\right)_{2},\ldots,\sigma).
\label{gpsi2}
\end{equation}

What would be the meaning of evolution equations like (\ref{gpsi}) and (\ref{gpsi2}), if they can be given any physical significance at all? First, it would be necessary to allow a Heisenberg uncertainty $\Delta g_{\mu\nu}$ to the metric at any given point $P_{l}$ in space-time. This uncertainty would correspond to the set of metrics at this point that cannot be excluded given an incomplete potential knowledge $PK_{g_{\mu\nu}}(n_{f})$ about the geometry of space-time just before the measurement of the metric within the context $C$ at sequential time $n_{f}$. When the measurement is made at time $n_{f}$, new knowledge about $g_{\mu\nu}$ is gained. From the present epistemic perspective this would mean that the geometry of space-time becomes more well-defined at point $P_{l}$ at a fundamental level.

In this sense, the above equations become expressions of quantum gravity. Just like different unobservable trajectories of an object may interfere constructively or destructively to produce a large or small probability to find it later at a given spatio-temporal position, different unobservable metrics might interfere in the determination of the probability to find a given metric at a later sequential time. Such interference would in turn affect the likelihood of trajectories of objects that move in the space-time described by this metric.

\begin{figure}[tp]
\begin{center}
\includegraphics[width=80mm,clip=true]{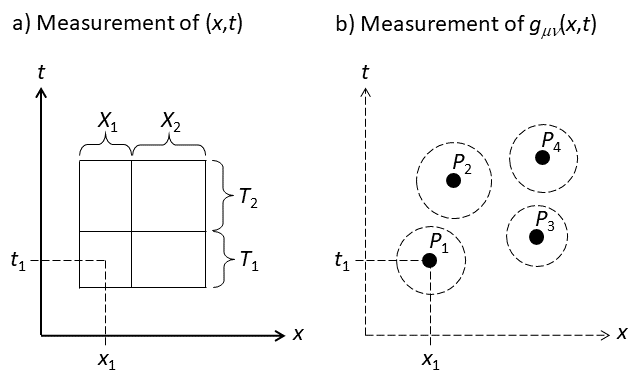}
\end{center}
\caption{(a) Measurement of the spatio-temporal location $(x,t)$ of some object $O_{l}$ gives a pair of alternatives $(X_{j},T_{j'})$ defined within a predetermined reference frame. Observers using a different reference frame to locate the same object arrives at a different result. (b) Measurement of $g_{\mu\nu}(x,t)$ gives $\left(G_{\mu\nu}\right)_{j}(P_{l})$ at some point $P_{l}$ defined by the location of a predetermined reference object $O_{l}$, where $\left(G{\mu\nu}\right)_{j}$ is an alternative among the set of possible metrics that could have been measured at $P_{l}$. A reference frame with coordinates $(x_{l},t_{l})$ can be introduced to specify this point, but it is not necessary, and different observers may use different coordinates to specify the same (transformed) metric at the same point $P_{l}$.}
\label{Figure5}
\end{figure}

In fact, the metric itself is numerically defined by means of spatio-temporal coordinates, and may be estimated via the measurement of trajectories of objects, as specified by a set of coordinates. Measurements of metrics and coordinates are intertwined. However, there is still a fundamental difference concerning the role of the coordinates in the two cases.

In a context $C$ in which $(x,t)$ is measured, it is measured for some object moving in space-time, and a coordinate system is presupposed as part of $PK_{C}(n_{i})$, as illustrated in Fig. \ref{Figure5}(a). In another context $C'$ in which $g_{\mu\nu}$ is measured, it is measured in relation to a reference object $O_{l}$ that defines the point $P_{l}$ at which the attribute is defined. $O_{l}$ must be known prior to the experiment, and its identity must be part of $PK_{C'}(n_{i})$. However, there does not need to be any predefined coordinate system. The relevant point $P_{l}$ may be identified with the same label $l$ as the corresponding reference object, as illustrated in Fig. \ref{Figure5}(b).

In the first case, the coordinates $(x,t)$ are clearly observables, but in the second case they become nothing more than labels that identify objects or points, for which some attribute other than the location itself is observed. Another set of labels may just as well be chosen, like a simple item number $l$ in a list. What is essential is the existence of a label that is shared among all observers to identify the object or point of interest. Referring to the discussion in section \ref{noconform}, it becomes evident that from the present cognitive perspective, the role of the coordinates depends on which attributes are measured in a context $C$, rather than on the chosen theoretical framework, be it general relativity or quantum theory. This fact is reflected in the different roles played by the coordinates in the contextually defined evolution equations (\ref{rtpsi}) and (\ref{gpsi}).

All experimental contexts $C$ have finite size and are localized around a set of observers. This fact limits the region of space-time that can be explored within such a context at any given time $n$, even though some measurements performed within $C$ may concern deduced attributes of faraway quasiobjects. Therefore, a local coordinate system centered around $C$ is always sufficient to describe the potential and actual outcome of the corresponding experiment.

The problem how to define a global coordinate system in general relativity therefore becomes irrelevant to the contextual evolution equations of the form (\ref{cspsi}) considered in this paper, even though the locally employed coordinate system must be applicable in an extended region around $C$, which sometimes reaches an astronomical or even a cosmological scale. In the same vein, the wave functions $\Psi$ appearing in these evolution equations are locally defined during the course of the experiment, in the sequential time interval $\left[n_{i},n_{f}\right]$. In other words, the physical meaning of the Hilbert space with the evolving state vector is considered to be the representation of a family of experimental contexts $C(\sigma)$. When an experiment in this family is concluded at time $n_{f}$, the corresponding wave function is no longer defined. Consequently, the 'universal Hilbert space' and the 'wave function of the universe' become meaningless concepts in the present approach. The fundamental state space is rather considered to be $\mathcal{S}$, in which the physical state $S$ evolves according to Eq. (\ref{evop}), as discussed in section \ref{cognitive}.

In contrast, Einstein's field equation (\ref{efield}) retains its global validity, of course, in cases where the Heisenberg uncertainty $\Delta g_{\mu\nu}$ of the metric can be neglected. Measurements within an experimental context $C$ of the metric around a series of spatio-temporal locations $\{P_{l}\}$ provide a sample that may help estimating the global geometry of space-time deductively from the field equation. Such measurements may reduce the number of space-times that cannot be excluded given the present potential knowledge. In other words, they may correspond to a state reduction as defined in Eq. (\ref{sored}), or a 'wave function collapse'. After a measurement that increases the knowledge about the geometry of space-time around a series of locations, the field equations may be solved anew, potentially increasing the knowledge about the global geometry, possibly even changing the expected topology. A local experiment may thus have global ramifications.

The consequences of applying a strict epistemic approach to space-time like the present one may be hard to swallow. Just as it allows the geometry of space-time to become more well-defined at a fundamental level by means of measurements, it allows the geometry to become less well-defined when the potential knowledge carried by all aware observers diminishes as a consequence of memory loss, or of a loss of records with a known correlation to the spatio-temporal geometry.

\section{Discussion}

This paper aims to show that the two aspects of time introduced and analysed in the accompanying paper \cite{ostborn} allow the formulation of generally covariant quantum mechanical evolution equations. These equations may describe the evolution of any observable quantity, for example the geometry of space-time. In so doing, they may be useful in the study of quantum gravity. It must be stressed, however, that the discussion in this paper is purely conceptual. The general form of the evolution equations is provided, but not the evolution operators that would allow quantitative predictions.

In a fortcoming paper, the evolution equation (\ref{rtpsi}) that applies when the spatio-temporal position of the specimen is measured will be analyzed in more detail. This equation corresponds to the Schr\"{o}dinger equation in ordinary quantum mechanics. In that equation, the Hamiltonian operator $\mathcal{H}$ evolves the state, and the energy eigenstates of $\mathcal{H}$ correspond to steady states with respect to $t$. Similarly, it will be argued that the operator $\mathcal{A}_{xt}$ in the evolution equation (\ref{rtpsi}) is proportional to the square of a rest energy operator with eigenstates that correspond to a specimen $OS$, as defined in section \ref{expcon}, with precisely defined rest mass. Such states will be steady states with respect to the evolution parameter $\sigma$. The fact that Lorentz transformations do not apply to the evolution parameter $\sigma$, as discussed in section \ref{empiricallaw}, is reflected by the fact that rest mass is a relativistic invariant. The corresponding steady state equations correspond to the Klein-Gordon or the Dirac equation.

No similar understanding has been reached yet by the author of the present paper of the nature of the evolution operator $\mathcal{A}_{g}$ in the evolution equation (\ref{gpsi}) for the space-time metric. Therefore, it cannot be claimed with any certainty that this equation is relevant to quantum gravity, and, if so, points at a solution to the problem of time.

To the author of this paper, the perceived severity of the incompatibility of general relativity and quantum theory stems from the idea that at least one of these theories expresses the fundamental nature of the physical world. Since some quantities in these two theories have different meaning and status, like time and the spatio-temporal coordinates, arguments arise about which of these two theories should take precedence of the other, leading to attempts to express the other theory in terms of the concepts and ideas of the preferred theory.

The idea that the preferred theoretical framework captures the essence of the physical world elevates the concepts and quantities used in that theory to the transcendental realm. In this spirit, some see the gravitational field expressed by the metric $g_{\mu\nu}$ as a substratum for the world, possibly together with the universal wave function or quantum fields. On top of these theoretical entities, everything else can be reconstructed, it is claimed, including scientists making experiments. 

The present epistemic approach turns this perspective on its head, in the spirit of Kant's Copernican revolution \cite{kant}. General relativity as well as quantum theory stem from our attempts to construct a quantitative model of the perceived world. One set of observations leads to relativity, another set leads to quantum theory. Rather than trying to resolve the apparent inconsistency between these two branches of the scientific endeavor by tying them together, or cutting one of them off, the idea explored here is to seek coherence at the common root - the fundamental forms of appearances that define the structure of the perceived world, such as time, and the basic structure of the experimental contexts that are necessary to gather and quantify appearances in a systematic manner. These cognitive structures are seen as the substratum on top of which scientific theories built.

At this cognitive root, two aspects of time were identified in the accompanying study \cite{ostborn}. Sequential time $n$ and the associated evolution parameter $\sigma$ can be associated to the universal time used to evolve states in Newtonian mechanics and quantum theory, whereas relational time $t$ can be associated to the temporal coordinate of space-time used in relativity. Both of them appear in the general form for quantum evolution equations that is motivated in section \ref{empiricallaw}.

Since these evolution equations are defined in specific families of experimental contexts, in which a given set of attributes of a specimen is measured, the domain of validity of these equations is limited to such contexts. In this picture, there is no wave function and no Hilbert space when there is no ongoing scientific experiment. Concepts like the wave function for the entire universe are dismissed as the result of confusion about what is more fundamental: the theoretical constructs or the cognitive preconditions for these constructs.

It is an eternal question whether theory is more fundamental than cognition, or vice versa. It was addressed almost 2 500 years ago by Democritus in \emph{a dialogue between the intellect and the senses}, according to Galen, fragment 125 \cite{lee}:

\begin{quote}
\emph{Intellect: Ostensibly there is colour, ostensibly sweetness, ostensibly bitterness, actually only atoms and the void.\\
Senses: Poor intellect, do you hope to defeat us while from us you borrow your evidence? Your victory is your defeat.}
\end{quote}
A modern counterpart to the intellect's vision of \emph{only atoms and the void} would be \emph{only gravitational and quantum fields}. 

From the present cognitive perspective, the role of the spatio-temporal coordinates is not determined once and for all by some preferred theoretical framework, but is given by the experimental context at hand. If the context is such that the position or the trajectory of the specimen is measured, the existence of a predefined coordinate system is crucial, of course. On the other hand, in a context where the metric or the curvature of space-time is measured, the coordinates become arbitrary labels used to identify the locations at which the values of these geometric attributes are determined, as illustrated in Fig. \ref{Figure5}.   

Likewise, the cognitive starting point dissolves the problem how to define a global coordinate system in general relativity, from which we can pick a universal temporal variable to use in generally covariant evolution equations. The evolution parameter employed here emerges from the cognitive analysis in the accompanying paper, and is distinct from the spatio-temporal coordinates. It provides a global flow of time, but lacks inherent metric. Spatio-temporal coordinates used in the proposed evolution equations can be considered local, since they are defined within a spatio-temporally localized context. Even though the specimen studied in this context may be far away in space and time, the assigned coordinates must remain finite whenever they are measured in a proper context with finite duration $n_{f}-n_{i}$. Therefore there is no need to define global coordinates from the empirical point of view, which is considered fundamental in this study.

\end{document}